\documentclass{JHEP3}

\newcommand{\e}{\begin{eqnarray}}
\newcommand{\ee}{\end{eqnarray}}

\newcommand{\CN}{{\cal N}}

\def\a{\alpha}

\def\d{\delta}

\newcommand{\p}{\psi}

\newcommand{\BZ}{\bar{Z}}
\title{Symplectic Three-Algebra and ${\cal N}=6, Sp(2N)\times U(1)$
Superconformal Chern-Simons-Matter Theory }

\author{Fa-Min Chen , Yong-Shi Wu \\
Department of Physics and Astronomy, University of Utah\\
Salt Lake City, UT 84112-0830, USA \\
E-mail:
\email{fchen@physics.utah.edu}, \email{wu@physics.utah.edu}}

\abstract{We introduce an {\em anti-symmetric} metric into a
3-algebra and call it a {\em symplectic} 3-algebra. The ${\cal
N}=6$, $Sp(2N)\times U(1)$  superconformal Chern-Simons-matter
theory with $SU(4)$ R-symmetry in three dimensions is constructed
by specifying the 3-brackets in a symplectic 3-algebra. We also
demonstrate that the $\CN=6$, $U(M)\times U(N)$ theory can be
recast into this symplectic 3-algebraic framework. }

\keywords{Symplectic Three-Algebra, Chern-Simons Theory, M2 branes}

\usepackage{bbm}
\usepackage{amsmath}
\usepackage{slashed}

\begin{document}

\section{Introduction} \label{Introduction}

Recently the construction of superconformal Chern-Simons-matter
(CSM) theories in three dimensions \cite{Schwarz2004,gaiottoyin} has
attracted a lot of attention in string/M theory community, because
they are natural candidates for the dual gauge description of M2
branes in M theory. It was already realized about twenty years ago
\cite{CSW1}, that generically Chern-Simons gauge theories in three
dimensions are conformally invariant, both for pure gauge theories
and for theories coupled to (massless) matter fields, even at the
quantum level: in spite of a quantum shift at one loop order
\cite{CSW0}, the Chern-Simons gauge coupling does not run at all,
because its $\beta$ function vanishes, as shown both by an explicit
two-loop calculations for theories with matter \cite{CSW2} and by a
formal proof up to all orders in perturbation theory for pure gauge
theories \cite{Piguet}. In order to construct the dual gauge
description of M2-branes, the relevant issue \cite{Schwarz2004} is
then how to incorporate {\em extended supersymmetries} into CSM
theories, since extended supersymmetry plays a crucial role in
M-theory as it does in superstring theory.

Based on the totally anti-symmetric Nambu 3-brackets
\cite{Nambu,Limiao:1999fm}, the maximally (i.e. ${\cal N}=8$)
supersymmetric Chern-Simons-matter theory in $D=3$ with $SO(4)$
gauge group and $SO(8)$ R-symmetry, was constructed independently by
Bagger, Lambert \cite{Bagger}, and Gustavsson \cite{Gustavsson}.
This theory, known as the BLG theory, is the dual gauge description
of two M2 branes \cite{DMPV,LambertTong}. It was also shown that the
Nambu 3-algebra with a symmetric and positive define metric is
unique \cite{Gauntlett, Papadopoulos}: it generates an $SO(4)$ gauge
symmetry. To generate arbitrary gauge group, the Nambu 3-algebras
with a Lorentzian metric (Lorentzian 3-algebra) are introduced
\cite{Lorentzian3Alg}. However, the BLG theory constructed from the
Lorentzian 3-algebras turns out to be an ${\cal N}=8$ super
Yang-Mills theory \cite{GhostFree,CalN8SYM}, which is not a
supersymmetric Chern-Simons-matter theory.

Soon the BLG theory was generalized to the cases of reduced
supersymmetries by Aharony, Bergman, Jafferis and Maldacena (ABJM)
\cite{ABJM}. They have been able to construct, without consulting to
the 3-algebra approach, an ${\cal N}=6$ superconformal CSM theory
with gauge group $U(N)\times U(N)$, $SU(4)$ R-symmetry and $U(1)$
global symmetry. They also argued that at level $k$, the theory
describes the low energy limit of $N$ M2-branes probing a
$\textbf{C}^4/\textbf{Z}_k$ singularity. At large-$N$ limit, it
becomes the dual theory of M theory on $AdS_4\times
S^7/\textbf{Z}_k$ \cite{ABJM}. Some further analysis of the ABJM
theory can be found in Ref. \cite{Benna,Schwarz}. The superconformal
gauge theories in $D=3$ with more or less supersymmetries can be
obtained by taking a conformal limit of $D=3$ gauged supergravity
theories \cite{Bergshoeff:2008cz,Bergshoeff}. By generalizing
Gaiotto and Witten's construction \cite{GaWi} of ${\cal N}=4$ CSM
theories, not only was the ABJM theory re-derived (as a special case
of $U(M)\times U(N)$ CSM theories), but also two new theories,
${\cal N}=5, Sp(2M)\times O(N)$ and ${\cal N}=6, Sp(2M)\times O(2)$
CSM theories, were constructed in Ref.
\cite{HosomichiJD,Hosomichi:2008jb}. Their M theory and string
theory dualities were studied in Ref. \cite{Aharony:2008gk}.

Bagger and Lambert \cite{Bagger08:3Alg} have constructed the
${\cal N}=6$, $U(M)\times U(N)$ theory in a modified 3-algebra
approach, in which the structure constants are no longer required
to be totally antisymmetric. By specifying the 3-brackets and
taking a hermitian, gauge invariant metric on the 3-algebra, they
are able to reproduce the $U(M)\times U(N)$ theory hence the ABJM
theory. However, it remains unclear whether another important
class of ${\cal N}=6$ CSM theories, namely the ones with gauge
group $Sp(2M)\times O(2)$, can be constructed in the 3-algebra
approach or not. In this paper we will propose to solve this
problem by introducing an \emph{anti-symmetric} metric into a
3-algebra, which we call a \emph{symplectic} 3-algebra. Then we
will first present a construction of ${\cal N}=6$ superconformal
CMS theories with $SU(4)$ R-symmetry by utilizing the symplectic
3-algebras, and then to re-derive the ${\cal N}=6,Sp(2N)\times
U(1)$ superconformal CSM theories by specifying the 3-bracket. We
will also demonstrate that the ${\cal N}=6$, $U(M)\times U(N)$
theory can be recast into this symplectic 3-algebraic framework.

The paper is organized as follows. In section \ref{3Alg}, we
introduce the notion of symplectic three-algebra. In section
\ref{ClsSUSY}, we study the supersymmetry transformations of
physical fields valued in a symplectic 3-algebra, and construct
the Lagrangian of the superconformal Chern-Simons-matter theories.
In section \ref{SymplecticSect}, we derive the ${\cal N}=6,
Sp(2N)\times U(1)$ superconformal CMS theory from our Lagrangian
by specifying the 3-brackets of the symplectic 3-algebra. In
section \ref{ugroup}, we recast the $\CN=6$, $U(M)\times U(N)$
theory into our framework. In section \ref{Conclusion}, we present
conclusions and discussions. Our convention and useful identities
are given in appendix \ref{Identities}.

\section{The Symplectic Three-Algebra}
\label{3Alg}

In this section, we will introduce the notion of the symplectic
3-algebra. Following \cite{Bagger08:3Alg}, we assume that the
3-algebra is a \emph{complex} vector space, equipped with the
3-brackets
\begin{eqnarray}\label{Symp3Bracket}
[T^I,T^J;T^K]=f^{IJK}{}_LT^{L},
\end{eqnarray}
where $T^I$ are a set of generators. ($I=1,2,\cdots,M$ ). Notice
that our 3-brackets are not exactly the same as that of Ref.
\cite{Bagger08:3Alg}. First, the third generator in the 3-brackets
is \emph{not} a complex conjugate one; secondly, we do \emph{not}
assume that the first two indices of the structure constants are
necessarily anti-symmetric.

Further we assume the structure constants satisfy the fundamental
identity (FI)
\begin{equation}
f^{IJK}{}_{O}f^{OLM}{}_N-f^{ILM}{}_{O}f^{OJK}{}_N+
f^{LJK}{}_{O}f^{IOM}{}_N +f^{MJK}{}_{O}f^{ILO}{}_N=0.
\label{FFI}
\end{equation}
In this way, the 3-algebra can be viewed as a generalization of the
ordinary (Lie) 2-algebra, with the FI playing the role of the Jacobi
identities. Note that our FI is also \emph{not} the same as that of
BL \cite{Bagger08:3Alg}. However, later we will see, one can `map'
the structure constants and the FI satisfied by the structure
constants in Ref. \cite{Bagger08:3Alg} into ours (see section
\ref{ugroup}).

Since our goal is to construct a theory with gauge group
$Sp(2N)\times U(1)$, it is natural to introduce an antisymmetric
metric $\omega^{IJ}$ and its inverse $\omega_{IJ}$ into the
3-algebra to raise or lower the indices:
\begin{eqnarray}\label{SymplecticStructure}
f^{IJKL}\equiv \omega^{LM}{f^{IJK}{}_{M}}, \quad f^{IJ}{}_{KL}\equiv
\omega_{KM}f^{IJM}{}_L,
\end{eqnarray}
where $\omega^{IJ}= -\omega^{IJ}$ and $\omega_{IJ}\omega^{JK}=
\delta_{I}{}^K$. The antisymmetric metric $\omega^{IJ}$ will enter
the theories via (\ref{SymplecticStructure}). We note that
the existence of the inverse of $\omega_{IJ}$ implies that ${\rm
det}(\omega_{IJ})\neq 0$, which in turn requires the complex
dimension $M$ of the 3-algebra be even $M=2L$.

Following BL \cite{Bagger08:3Alg}, we define the global
transformation of a 3-algebra valued field $X_I$ as
\begin{equation}\label{GlobalSym}
\delta_{\tilde\Lambda}
X_I=-\Lambda^K{}_Lf^{JL}{}_{KI}X_J\equiv-\tilde\Lambda^J{}_I X_J,
\end{equation}
where the parameter $\Lambda^K{}_L$ is a 3-algebra tensor,
independent of spacetime coordinate. The anti-symmetric metric must
be invariant under such a global transformation:
\begin{eqnarray}\label{ConstrOnf}\nonumber
\delta_{\tilde\Lambda}\omega_{IJ}&=&-\tilde{\Lambda}^K{}_I\omega_{KJ}
-\tilde{\Lambda}^K{}_J\omega_{IK}\\
&=&-\Lambda^L{}_M(f^{KM}{}_{LI}\omega_{KJ}+f^{KM}{}_{LJ}\omega_{IK})\\
\nonumber &=&0.
\end{eqnarray}
From point of view of ordinary Lie group, the infinitesimal matrices
$-\tilde{\Lambda}^K{}_I$ must form the Lie algebra
$Sp(2L,\mathbb{C})$. Parts of the global symmetry (\ref{GlobalSym})
will be gauged in section \ref{ClsSUSY} and \ref{ugroup}. Eq.
(\ref{ConstrOnf}) imposes a strong constraint on $f^{IJ}{}_{KL}$.
The FI (\ref{FFI}) implies that the structure constants are also
invariant under the global transformation \cite{Bagger08:3Alg}:
\begin{eqnarray}\label{FI5}
\delta_{\tilde\Lambda}f^{IJ}{}_{KL}=0
\end{eqnarray}

We call the 3-algebra defined by the antisymmetric metric and the
above Eqs. (\ref{Symp3Bracket})-(\ref{SymplecticStructure}) a
symplectic 3-algebra.

Moreover, since the 3-algebra is a complex vector space, the reality
conditions and positivity are important in introducing gauge fields,
matter fields and constructing invariant Lagrangians. Following BL
\cite{Bagger08:3Alg}, we define the gauge fields as
\begin{eqnarray}\label{DefGauge}
\tilde{A}_\mu{}^I{}_L\equiv A_\mu{}^K{}_Jf^{IJ}{}_{KL},
\end{eqnarray}
where $A_\mu{}^K{}_J$ is a 3-algebra tensor ($\mu=0, 1, 2$), satisfying
the anti-hermitian condition
\begin{eqnarray}\label{RealG}
A_\mu{}^{*K}{}_J=-A_\mu{}^{J}{}_K.
\end{eqnarray}
To ensure the anti-hermiticity of the gauge field, i.e.,
$\tilde{A}_\mu{}^{*I}{}_L=-\tilde{A}_\mu{}^{L}{}_I$, the structure
constants $f^{IJ}{}_{KL}$ are required to satisfy the reality
condition
\begin{eqnarray}\label{HermiCondi}
f^{*IJ}{}_{KL} =f^{LK}{}_{JI }.
\end{eqnarray}

In accordance with (\ref{RealG}), we also require that the parameter
in (\ref{GlobalSym})
satisfies the anti-hermitian condition
$\Lambda^{*K}{}_J=-\Lambda^{J}{}_K$.
On the other hand, it obeys the natural reality condition
$\Lambda^{*K}{}_J=-\omega_{KI}\omega^{JL}\Lambda^{I}{}_L$,
since it carries two symplectic 3-algebra indices.
These two equations imply that $\Lambda_{KJ}=\Lambda_{JK}$.
Re-examining the global transformation (\ref{GlobalSym}),
we are led to require that the structure constants are symmetric
in the middle two indices
\begin{eqnarray}{\label{SymofF}}
f_{IJKL}=f_{IKLJ}.
\end{eqnarray}

The reality conditions for the matter fields are a bit more
complicated, since they involve additional indices associated with
the $R$-symmetry, which will be discussed in next section. As for
the positivity of invariant Lagrangians, the following hermitian
bilinear form in the 3-algebra (as a complex vector space) is
naturally positive-definite:
\begin{eqnarray}\label{HermiInnerProdu}
h(X,Y) = X^*_I Y_I,
\end{eqnarray}
where $*$ is the complex conjugation. Using it to construct the
Lagrangians in our model will guarantee their positivity. But are
they invariant under the transformations that preserve the
anti-symmetric metric? Generally this is not true. But fortunately
it is known that $Sp(2L, \mathbb{C})$ and $U(2L)$ have a non-empty
intersection $Sp(2L)$, which can be selected by imposing certain
reality conditions on the fields. In fact, eqn. (\ref{GlobalSym})
and reality condition (\ref{HermiCondi}) dictate the transformation
property of $X^*_I\equiv \bar X^I$ to be $\delta_{\tilde\Lambda}\bar
X^I=\tilde{\Lambda}^I{}_J\bar X^J$:
\begin{eqnarray}\nonumber
\delta_{\tilde\Lambda} X^*_I&=&-\Lambda^{*K}{}_Lf^{*JL}{}_{KI}X^*_J\\
\nonumber
&=&-(-\Lambda^L{}_K)f^{IK}{}_{LJ}X^*_J\\
&=&\tilde{\Lambda}^I{}_JX^*_J . \label{GaugeTransfOfZstar}
\end{eqnarray}
The bilinear form (\ref{HermiInnerProdu}) is therefore compatible
with the antisymmetric metric $\omega^{IJ}$, in the sense that with
the reality conditions respected, the complex conjugate $X^*_{I}$
transforms in the same way as a vector $\omega^{IJ} X_J$.
Essentially this means that while the reality conditions are
respected, it makes sense to denote $X^*_{I}$ as $\bar{X}^I$, and
rewrite eq. (\ref{HermiInnerProdu}) in a manifestly invariant form:
$h(X,Y) = \bar{X}^I Y_I$. Therefore the terms in the action
constructed in terms of the hermitian bilinear form will be $Sp(2L)$
invariant.

\section{${\cal N}=6, Sp(2N)\times U(1)$ CSM Theory from Symplectic 3-Algebra}
\label{3AlgCMS}

\subsection{Closure of the Super-algebra}\label{ClsSUSY}
In this subsection, we generalize the method of BL's 3-algebra
construction of ${\cal N}=6$ CSM theories \cite{Bagger08:3Alg}. The
formalism and computations are similar to theirs except for
necessary changes arising from the fact that we introduce an
antisymmetric metric $\omega_{IJ}$ into the 3-algebra and hence the
theories.

We first postulate that scalar and fermion fields are symplectic
3-algebra valued, carrying a vector index, while the gauge fields
are defined by (\ref{DefGauge}).

To generate a direct product of gauge group, such as $Sp(2N)\times
U(1)$, we split up an 3-algebra index into a pair of indices
$I\rightarrow a\pm$, where the index $I$ runs from 1 to 4N, while
the index $a$ from 1 to 2N. The index $a$ is an $Sp(2N)$ index.
And $+$ or $-$ is a $U(1)=SO(2)$ index.

We then assume the theory has an $SU(4)$ R-symmetry and a $U(1)$
global symmetry. Combining the $SU(4)$ R-symmetry and $U(1)$ global
symmetry, the complex scalar fields can be written as $Z^A_{c+}$
($A=1,2,3,4$), and their corresponding complex conjugates
$\bar{Z}^{c+}_{A}\equiv Z^{*A}_{c+}$, where $A$ is the R-symmetry
index. (The $U(1)$ index $+$ will be suppressed later.) We will
follow a convention similar to that of Bagger and Lambert
\cite{Bagger08:3Alg}: a superscript $A$ indicates the fundamental
representation $\bf 4$ of $SU(4)$; a subscript $A$ indicates the
$\bar{\bf 4}$ of $SU(4)$. Similarly, we label the fermionic fields
by $\psi_{Ac+}$ and $\psi^{Ac+}$. We also assign a unit global
$U(1)$ charge to $Z^A_{c+}$ and $\psi_{Ac+}$.  The definition of the
matter fields suggests that the parameter  $\Lambda^K{}_J$ (see
(\ref{GlobalSym})) takes the form
\begin{eqnarray}\label{GaugePrmt}
\Lambda^K{}_J=\Lambda^{c+}{}_{b+}.
\end{eqnarray}
We will see that (\ref{GaugePrmt}) is reasonable when we examine
the closure of the super-algebra. (Actually, if we restore the
$U(1)$ index $+$ in Eq. (\ref{GaugeTransf}), i.e. replace $b$ and
$c$ by $b+$ and $c+$, respectively, the parameter $\Lambda$ will
exactly have the above form.) Similarly, the 3-algebra tensor
$A_\mu{}^K{}_J$ takes the form $A_\mu{}^{c+}{}_{b+}$. So the gauge
fields are given by $\tilde
A_\mu{}^{a+}{}_{d+}=A_\mu{}^{c+}{}_{b+}f^{a+b+}{}_{c+d+}$.

We denote the antisymmetric metric $\omega^{IJ}$ as
\begin{eqnarray}\label{AntiMetr}
\omega^{a+,b-}\equiv \omega^{ab}h^{+-},
\end{eqnarray}
where $\omega^{ab}$ is an antisymmetric
bilinear form, and $h^{+-}=h^{-+}=1$. The other components of the antisymmetric metric $\omega^{IJ}$ vanish due to the fact that $h^{++}=h^{--}=0$.

The hermitian bilinear form (\ref{HermiInnerProdu}) becomes
\begin{eqnarray}\label{HermiInnerProdu2}
h(X,Y)=\bar{X}^{a+}Y_{a+}.
\end{eqnarray}

Because of the index structure of $\Lambda^{c+}{}_{b+}$, the two
types of gauge parameters can be written as
$\tilde{\Lambda}^{a+}{}_{d+}=\Lambda{}^{c+}{}_{b+}f^{a+b+}{}_{c+d+}$
and
$\tilde{\Lambda}^{a-}{}_{d-}=\Lambda{}^{c+}{}_{b+}f^{a-b+}{}_{c+d-}$.
At first it seems that we need both $f^{a+b+}{}_{c+d+}$ and
$f^{a-b+}{}_{c+d-}$ to construct a gauge theory. But the gauge
invariance condition of $\omega$ (\ref{ConstrOnf}) implies that
\begin{eqnarray}\label{SymmeOfF3}
f_{a-b-c+d+}=f_{d+b-c+a-},
\end{eqnarray}
so $f^{a+b+}{}_{c+d+}$ alone will be sufficient to construct a gauge
theory. As a result, we only need to consider the FI satisfied by
$f^{a+b+}{}_{c+d+}$, or the gauge invariance condition
\begin{eqnarray}\label{FI6}
\delta_{\tilde{\Lambda}}f^{a+b+}{}_{c+d+}=0
\end{eqnarray} in (\ref{FI5}).

Eq. (\ref{SymofF}) can be written as
\begin{eqnarray}\label{SymmeOfF4}
f_{a-b-c+d+}=f_{a-c+b-d+}.
\end{eqnarray}

The reality condition (\ref{HermiCondi}) becomes
\begin{eqnarray}\label{HermiCondi2}
f^{*a+b+}{}_{c+d+}=f^{d+c+}{}_{b+c+}.
\end{eqnarray}

To close the ${\cal N}=6$ super-algebra, we need to impose an additional
 constraint condition on the structure constants. Specifically, we
will require the first two indices of the structure constants are
anti-symmetric if they have the same gauge transformation property,
or if they are on equal footing, i.e.
\begin{equation}\label{SymmeOfF}
f_{a-b-c+d+}=-f_{b-a-c+d+}.
\end{equation}
However, the first two pairs of $f_{d+b-c+a-}$ (see (\ref{SymmeOfF3}))
are \emph{not} necessarily anti-symmetric, since $d+$ and $b-$
do \emph{not} have the same gauge transformation property.

We now suppress the $SO(2)$ indices for the sake of brevity. We
write the structure constants $f^{a+b+}{}_{c+d+}$ as
$f^{ab}{}_{cd}$. Similarly, we write the scalar, fermion, gauge
fields and the gauge parameter as $Z^A_{c}, \psi^{Ac}$, $\tilde
A_\mu{}^{a}{}_{d}=A_\mu{}^{c}{}_{b}f^{ab}{}_{cd}$ and
$\Lambda{}^{c}{}_{b}$, respectively. The hermitian bilinear form
(\ref{HermiInnerProdu2}) can be written as $\bar{X}^{a}Y_{a}$. Using
the anti-symmetric metric (\ref{AntiMetr}) to raise $a-$ and $b-$ in
Eq. (\ref{SymmeOfF}), then Eq. (\ref{SymmeOfF}) becomes
\begin{eqnarray}\label{SymmeOfF5}
f^{ab}{}_{cd}=-f^{ba}{}_{cd}.
\end{eqnarray}
Now the reality condition (\ref{HermiCondi2}) takes the following form:
\begin{eqnarray}\label{HermiCondi3}
f^{*ab}{}_{cd}=f^{dc}{}_{ba}=f^{cd}{}_{ab}.
\end{eqnarray}
And we write (\ref{FI6}) as  $\delta_{\tilde{\Lambda}}f^{ab}{}_{cd}=0$, which is equivalent to
\begin{equation}\label{FI}
f^{ab}{}_{cd}f^{de}{}_{gf} +f^{ba}{}_{fd}f^{de}{}_{gc}-
f^{ae}{}_{gd}f^{db}{}_{cf} -f^{be}{}_{gd}f^{da}{}_{fc}=0.
\end{equation}
Later we will use this form of the FI's in proving the closure of
the supersymmetry algebras. The FI (\ref{FI}) may also be written in some other forms; for
example,
\begin{equation}
f^{ab}{}_{cd}f^{de}{}_{fg} -f^{ae}{}_{fd}f^{db}{}_{cg}+
f^{eb}{}_{cd}f^{da}{}_{gf} -f^{ea}{}_{gd}f^{db}{}_{cf}=0.
\label{FI1}
\end{equation}

For ${\cal N}=6$ SUSY, the supersymmetry parameters are in the
fundamental representation of $SO(6)$: $\epsilon^I$, $I=1,...,6.$
Since $SO(6)\cong SU(4)$, we can relabel these generators by
two $SU(4)$ indices: $\epsilon^{AB}=-\epsilon^{BA}$. Namely,
they transform as the $\bf 6$ of $SU(4)$. The reality
condition $\epsilon^{*}_{AB}=\epsilon^{AB} = \frac{1}{2}
\varepsilon^{ABCD}\epsilon_{CD}$ implies that they do not
carry a global $U(1)$ charge.

To achieve conformal invariance, we assume that the local field
theory is scale invariant. Under this assumption, we then propose
the following manifest $SU(4)$ R-symmetry, ${\cal N}=6$ SUSY
transformations \footnote [1]{Our $f^{ab}{}_{cd}$ are actually
$f^{a+b+}{}_{c+d+}=\omega_{c+,e-}f^{a+b+e-}{}_{d+}$. If one replaces
our $f^{ab}{}_{cd}$ with BL's $f^{ab\bar c}{}_d$ , many equations in
this section take the same forms as those of BL's
\cite{Bagger08:3Alg}. 
}:
\begin{eqnarray}\label{susy}
\nonumber  \delta Z^A_d &=& i\bar\epsilon^{AB}\psi_{Bd} \\
 \nonumber
 \delta \bar{Z}_{A}^{d}
&=& i\bar\epsilon_{AB}\psi^{Bd} \\
 \nonumber
\delta \psi_{Bd} &=& \gamma^\mu D_\mu Z^A_d\epsilon_{AB} +
  f^{ab}{}_{cd}Z^C_aZ^A_b \bar{Z}_{C}^{c} \epsilon_{AB}+f^{ab}{}_{cd}
  Z^C_a Z^D_{b} \bar{Z}_{B}^{c}\epsilon_{CD} \\
\delta \psi^{Bd} &=& \gamma^\mu D_\mu \bar{Z}_A^d\epsilon^{AB} +
  f^{cd}{}_{ab}\bar{Z}_C^a \bar{Z}_A^b Z^{C}_{c} \epsilon^{AB}
+f^{cd}{}_{ab}\bar{Z}_C^a \bar{Z}_D^{b} Z^{B}_{c}\epsilon^{CD} \\
\nonumber
 \delta \tilde A_\mu{}^c{}_d &=&
-i\bar\epsilon_{AB}\gamma_\mu Z^A_a\psi^{Bb} f^{ca}{}_{bd} +
i\bar\epsilon^{AB}\gamma_\mu \bar{Z}_{A}^{a}\psi_{Bb}f^{cb}{}_{ad}.
\end{eqnarray}

The covariant derivatives are defined as
\begin{eqnarray}
D_\mu Z^A_d &=&
\partial_\mu Z^A_d -\tilde A_\mu{}^c{}_dZ^A_c\\
D_\mu \bar{Z}_A^d &=&
\partial_\mu \bar{Z}_A^d +\tilde A_\mu{}^d{}_c\bar{Z}_A^c,
\end{eqnarray}
and similar expressions for the fermionic fields.

We require the on-shell closure of the supersymmetry algebra.
Namely, after imposing equations of motion, the commutator of two
supersymmetry transformations must be equal to a translation plus a
gauge term.

The commutator of two supersymmetry transformations acting on the
scalar fields reads
\begin{eqnarray}
[\delta_1,\delta_2]Z^A_d = v^\mu \partial_\mu Z^A_d +
(\tilde{\Lambda}^a{}_{d}-v^{\mu}\tilde{A}_{\mu}{}^a{}_{d})Z^A_a,
\label{CloseOnScalars}
\end{eqnarray}
where
\begin{eqnarray}
v^\mu &=& \frac{i}{2}\bar\epsilon_2^{CD}\gamma^\mu\epsilon_{1CD},\\
\tilde{\Lambda}^a{}_{d}&=&\Lambda^c{}_bf^{ab}{}_{cd}\\
\Lambda^c{}_b&=&i(\bar{\epsilon}^{DE}_2\epsilon_{1CE}
-\bar{\epsilon}^{DE}_1\epsilon_{2CE})\bar{Z}_D^cZ^C_b\label{GaugeTransf}
\end{eqnarray}
The first term of eq. (\ref{CloseOnScalars}) is a translation, and
the second represents a gauge transformation, as expected. In
deriving (\ref{CloseOnScalars}), we have used Eq.
(\ref{SymmeOfF5}): $f^{ab}{}_{cd}=-f^{ba}{}_{cd}$.

For the gauge field, using the FI (\ref{FI}) and some identities in
the Appendix, we obtain
\begin{eqnarray}\label{Aclose}
[\delta_1,\delta_2]\tilde A_\mu{}^c{}_d &=&
v^{\nu}\partial_{\nu}\tilde A_\mu{}^c{}_d+
D_{\mu}(\tilde{\Lambda}^c{}_{d}-v^{\nu}\tilde{A}_{\nu}{}^c{}_{d})\\
\nonumber &&+v^{\nu}\bigg[\tilde{F}_{\mu\nu}{}^c{}_d
+\varepsilon_{\mu\nu\lambda}\left(D^\lambda Z^A_a \bar{Z}_A^b-
Z^A_aD^\lambda \bar{Z}_A^b
-i\bar\psi^{Ab}\gamma^\lambda\psi_{Aa}\right)f^{ac}{}_{bd}\bigg].
\end{eqnarray}
where $\tilde F_{\mu\nu}{}^c{}_d = \partial_\mu \tilde
A_\nu{}^c{}_d-\partial_\nu \tilde A_\mu{}^c{}_d+[\tilde A_\mu,\tilde
A_\nu]{}^c{}_d$ is the field strength. We recognize the first term
as a translation, and the second a gauge transformation. To achieve
the closure, we need to impose the following equation of motion for
the gauge field:
\begin{eqnarray}\label{EOMofGauge}
\tilde{F}_{\mu\nu}{}^c{}_d
=-\varepsilon_{\mu\nu\lambda}\left(D^\lambda Z^A_a \bar{Z}_A^b-
Z^A_aD^\lambda \bar{Z}_A^b
-i\bar\psi^{Ab}\gamma^\lambda\psi_{Aa}\right)f^{ac}{}_{bd}.
\end{eqnarray}
As BL discovered \cite{Bagger08:3Alg}, the FI implies
$D_{\mu}f^{ca}{}_{bd}=0$, if one writes $\tilde A_\mu{}^c{}_d=
A_\mu{}^b{}_af^{ca}{}_{bd}$ in the expression of the covariant
derivative. We have used this important equation to derive the
second term in eq. (\ref{Aclose}):
$f^{ca}{}_{bd}D_{\mu}\Lambda^b{}_{a}=D_{\mu}\tilde{\Lambda}^c{}_{d}$.

The commutator of two supersymmetry transformations acting on the
fermionic fields reads
\begin{eqnarray}
\nonumber [\delta_1,\delta_2]\psi_{Dd} &=& v^\mu \partial_\mu
\psi_{Dd} + (\tilde{\Lambda}^a{}_{d}-v^{\mu}\tilde{A}_{\mu}{}^a{}_{d})
\psi_{Da}\\
\nonumber &&-\frac{i}{2}(\bar\epsilon_1^{AC}\epsilon_{2AD}
-\bar\epsilon_2^{AC}\epsilon_{1AD})E_{Cd}\\
 &&
+\frac{i}{4}(\bar\epsilon^{AB}_1\gamma_\nu\epsilon_{2AB})\gamma^\nu E_{Dd},
\end{eqnarray}
where
\begin{equation}
E_{Cd} = \gamma^\mu D_\mu\psi_{Cd} +f^{ab}{}_{cd} \bigg(\psi_{Ca}
Z^D_b\bar{Z}_{D}^c-2\psi_{Da}Z^D_b\bar{Z}_{C}^c-\varepsilon_{CDEF}
\psi^{Dc}Z^E_aZ^F_b\bigg).
\end{equation}
Again, the first two term are a translation and a gauge
transformation, respectively. To achieve the closure of the
supersymmetry algebra, we have to impose the following equations of
motion for the fermionic fields:
\begin{equation}\label{EOMofFermion}
0=E_{Cd} = \gamma^\mu D_\mu\psi_{Cd} +f^{ab}{}_{cd} \bigg(\psi_{Ca}
Z^D_b\bar{Z}_{D}^c-2\psi_{Da}Z^D_b\bar{Z}_{C}^c
-\varepsilon_{CDEF}\psi^{Dc}Z^E_aZ^F_b\bigg).
\end{equation}

To derive the equations of motion of the scalar fields, we take
the super-variation of the equations of motion of the fermionic
fields: $\delta E_{Cd}=0$. Two equations are obtained: One is
\begin{eqnarray}\label{EOMofScalar}
&&0=D_{\mu}D^{\mu}Z^B_c-if^{ab}{}_{cd}
\bigg(\bar{\psi}^{Ad}\psi_{Aa}Z^B_b -2\bar{\psi}^{Bd}\psi_{Aa}Z^A_b
-\varepsilon^{ABCD}\bar{\psi}_{Aa}\psi_{Cb}\bar{Z}_D^d\bigg)\\
\nonumber
&&+\frac{1}{3}\bigg(f^{ae}{}_{fd}f^{bd}{}_{cg}
-2f^{ab}{}_{cd}f^{ed}{}_{fg} -2f^{db}{}_{gc}f^{ae}{}_{fd}
+2f^{ab}{}_{fd}f^{ed}{}_{cg}-4f^{eb}{}_{fd}f^{ad}{}_{cg}\bigg)
Z^B_e\bar{Z}^f_AZ^A_a\bar{Z}^g_DZ^D_b.
\end{eqnarray}
The other equation is equivalent to the equation of motion of the
gauge field (\ref{EOMofGauge}).

The equations of motions of the gauge, fermion and scalar fields,
Eqs. (\ref{EOMofGauge}), (\ref{EOMofFermion}) and
(\ref{EOMofScalar}) respectively, can be derived from the following
Lagrangian:
\begin{eqnarray}\label{Lagrangian}
\nonumber {\cal L} &=& -D_\mu \bar{Z}_A^aD^\mu Z^A_a -
i\bar\psi^{Aa}\gamma^\mu D_\mu\psi_{Aa} -V-{\cal L}_{CS}\\
&& -if^{ab}{}_{cd}\bar\psi^{Ad} \psi_{Aa}
Z^B_b\bar{Z}_B^c+2if^{ab}{}_{cd}\bar\psi^{Ad}
\psi_{Ba}Z^B_b\bar{Z}_A^c\\
\nonumber
&&-\frac{i}{2}\varepsilon_{ABCD}f^{ab}{}_{cd}\bar\psi^{Ac}
\psi^{Bd}Z^C_aZ^D_b
-\frac{i}{2}\varepsilon^{ABCD}f^{cd}{}_{ab}
\bar\psi_{Ac}\psi_{Bd}\bar{Z}_C^a\bar{Z}_D^b .
\end{eqnarray}
Here the scalar potential $V$ is
\begin{eqnarray}\label{Potential}
V=\frac{2}{3}\bigg(f^{ab}{}_{cd}f^{ed}{}_{fg}
-\frac{1}{2}f^{eb}{}_{cd}f^{ad}{}_{fg}\bigg)
\bar{Z}_A^c Z^A_e\bar{Z}_B^f Z^B_a\bar{Z}_D^g Z^D_b .
\end{eqnarray}
It can be recast into the following form \cite{Bagger08:3Alg}:
\begin{equation}
V = \frac{2}{3}\Upsilon^{CD}_{Bd}\bar\Upsilon_{CD}^{Bd} ,
\end{equation}
where
\begin{equation}\label{Upsilon}
\Upsilon^{CD}_{Bd} = f^{ab}{}_{cd}\bigg(Z^C_aZ^D_b\bar{Z}_B^c
-\frac{1}{2}\delta^C_BZ^E_aZ^D_b\bar{Z}_E^c+\frac{1}{2}
\delta^D_BZ^E_aZ^C_b\bar{Z}_E^c\bigg),
\end{equation}
and the quantity $\bar\Upsilon_{CD}^{Bd}$ is the complex conjugate
of $\Upsilon^{CD}_{Bd}$:
\begin{eqnarray}\nonumber
\bar\Upsilon_{CD}^{Bd}&=&\Upsilon^{*CD}_{Bd}\\
 &=& f^{cd}{}_{ab}\bigg(\bar{Z}_C^a\bar{Z}_D^b Z^B_c
-\frac{1}{2}\delta_C^B\bar{Z}_E^a\bar{Z}_D^b Z^E_c+\frac{1}{2}
\delta_D^B\bar{Z}_E^a\bar{Z}_C^bZ^E_c\bigg),
\label{UpsilonConjuga}
\end{eqnarray}
where the reality condition of the structure constants
$f^{*ab}{}_{cd}=f^{cd}{}_{ab}$ has been used.

The Chern-Simons term in the Lagrangian is
\begin{equation}
{\cal L}_{CS}=\frac{1}{2}\varepsilon^{\mu\nu\lambda}
(f^{ab}{}_{cd}A_{\mu}{}^c{}_b\partial_\nu A_{\lambda}{}^d{}_a
+\frac{2}{3}f^{ac}{}_{dg}f^{ge}{}_{fb}
A_{\mu}{}^b{}_aA_{\nu}{}^d{}_c A_{\lambda}{}^f{}_e).
\end{equation}

Notice that we have used the (positive definite) hermitian bilinear
form (\ref{HermiInnerProdu}) to construct the Lagrangian of the
matter fields. This bilinear form is invariant under our gauge
transformations which preserve the symplectic metric. Therefore our
Lagrangian is gauge invariant. To derive the equations of motion of
the scalar from the Lagrangian, one needs to use the FI (\ref{FI}).
The equations of motion derived from the Lagrangian
(\ref{Lagrangian}) are invariant under the 12 super-symmetries, and
reproduce those we have imposed before for on-shell closure of the
super-symmetries.


\subsection{${\cal N}=6, Sp(2N)\times U(1)$ CSM Theory}
\label{SymplecticSect}

We first specify the structure constants as
\begin{eqnarray}\label{SyplStruc1}
f_{a-,b-,c+,d+} = -k[(\omega_{ab}\omega_{cd}
+\omega_{ac}\omega_{bd})h_{-+}h_{-+}
+(\omega_{ad}\epsilon_{-+})(\omega_{bc}\epsilon_{-+})].
\end{eqnarray}
where $h_{+-}=h_{-+}=1$ and $\epsilon_{+-}=-\epsilon_{-+}=ih_{+-}$.
The structure constants have the symmetry property
(\ref{SymmeOfF3}), (\ref{SymmeOfF4}) and (\ref{SymmeOfF}), obey
the reality condition (\ref{HermiCondi2}), and satisfy the FI or (\ref{FI6}).
We observe that $f_{d+,b-,c+,a-}\neq f_{b-,d+,c+,a-}=0$. However,
this is not inconsistent with eq. (\ref{SymmeOfF}), since
$d+$ and $b-$ do \emph{not} have the same gauge transformation property.
Using $\omega^{a+,b-}$ to raise the first two pairs of indices of the
structure constants, we get

\begin{eqnarray}\label{SyplStruc3}
f^{a+b+}{}_{c+d+} = k[(\omega^{ab}\omega_{cd}
-\delta^a{}_c\delta^b{}_d)\delta^{+}{}_{+} \delta^{+}{}_{+} -
(\delta^a{}_d)(-i
\delta^{+}{}_{+})(\delta^b{}_c)(-i\delta^{+}{}_{+})],
\end{eqnarray}
where we have used $\epsilon_{-+}=-ih_{-+}$. Suppressing the $SO(2)$
indices gives
\begin{eqnarray}\label{SyplStruc2}
f^{ab}{}_{cd} = k(\omega^{ab}\omega_{cd}-\delta^a{}_c\delta^b{}_d
+\delta^a{}_d\delta^b{}_c).
\end{eqnarray}
We notice that (\ref{SyplStruc2}) takes the same form as the components
of an embedding tensor in Ref. \cite{Bergshoeff}. This is not just an
accident, and we will investigate their connection further in a coming
paper.

Substituting eq. (\ref{SyplStruc2}) into eq. (\ref{Lagrangian}), and
replacing $ A_\mu{}^b{}_a$ in the Lagrangian by $\frac{1}{k}
A_\mu{}^b{}_a$, we obtain the following Lagrangian:
\begin{eqnarray}
\label{SymplLagrangian}
\nonumber {\cal L} &=& -D_\mu \bar{Z}_A^aD^\mu Z^A_a -
i\bar\psi^{Aa}\gamma^\mu D_\mu\psi_{Aa} -V-{\cal L}_{CS}\\
&& +ik\bigg(\bar{Z}^b_B\bar \psi_{Ab}\psi^{Aa}Z^B_a-\bar Z_B^b
Z^B_b\bar\psi^{Aa}\psi_{Aa}-\bar
Z_B^c\omega_{cd}\bar\psi^{Ad}\psi_{Aa}\omega^{ab}Z^B_b\bigg)\\
\nonumber
&& -2ik\bigg(\bar{Z}^b_B\bar \psi_{Ab}\psi^{Ba}Z^A_a-\bar Z_B^b
Z^A_b\bar\psi^{Ba}\psi_{Aa}-\bar
Z_B^c\omega_{cd}\bar\psi^{Bd}\psi_{Aa}\omega^{ab}Z^A_b\bigg)\\
\nonumber
&& -ik\varepsilon^{ABCD}\bigg(\bar{Z}^a_A\bar \psi_{Ba}\bar
Z_C^b\psi_{Db}-\frac{1}{2}\bar Z_A^c\omega_{cd}\bar
Z_C^d\bar\psi_{Ba}\omega^{ab}\psi_{Db}\bigg)\\
\nonumber
&&-ik\varepsilon_{ABCD}\bigg(\bar\psi^{Ba}Z_a^A\psi^{Db}Z^C_b
-\frac{1}{2}Z^A_a\omega^{ab} Z^C_b\bar\psi^{Bc}
\omega_{cd}\psi^{Dd}\bigg).
\end{eqnarray}
Here the potential is
\begin{eqnarray}\label{SympPotential}
-V&=&-3k^2Z^B_a\omega^{ab}Z^D_b\bar Z_D^eZ^A_e \bar
Z_A^c\omega_{cd}\bar Z_B^d
+\frac{5k^2}{3}\bar Z_A^aZ^B_a \bar Z_B^bZ^D_b\bar Z_D^cZ^A_c\\
\nonumber &&-2k^2\bar Z_A^aZ^B_a\bar Z_D^bZ^D_b\bar Z_B^c Z^A_c
+\frac{k^2}{3}\bar Z_B^aZ^B_a\bar Z_D^bZ^D_b\bar Z_A^c Z^A_c,
\end{eqnarray}
and the Chern-Simons term is
\begin{eqnarray}
{\cal
L}_{CS}=\frac{1}{2k}\varepsilon^{\mu\nu\lambda}A_\mu\partial_\nu
A_\lambda-\frac{1}{4k}\varepsilon^{\mu\nu\lambda}
tr\bigg(B_\mu\partial_\nu B_\lambda
+\frac{2}{3}B_\mu B_\nu B_\lambda\bigg),
\end{eqnarray}
where the gauge fields $B_\mu{}^c{}_d\equiv -(A_{\mu d}{}^{c}+A_\mu{}^c{}_d)$ and
$A_\mu\equiv A_\mu{}^a{}_a$ are defined by the following equation:
\begin{eqnarray}
\tilde A_\mu{}^c{}_d &=& A_\mu{}^b{}_af^{ca}{}_{bd}\\
\nonumber
&=&-(A_{\mu d}{}^{c}+A_\mu{}^c{}_d)+(A_\mu{}^a{}_a)\delta^c{}_d.
\end{eqnarray}
The expression of the structure constants (\ref{SyplStruc2}) has
been used. We recognize that $A_\mu$ is the $U(1)$ part of the gauge
potential, and $B_\mu{}^c{}_d$ the $Sp(2N)$ part, since it can be
written as $B_\mu{}^c{}_d=A_\mu^{ab}(t_{ab})^c{}_d$, where
$(t_{ab})^c{}_d$ is in the defining representation of the Lie
2-algebra of $Sp(2N)$.

The Lie 2-algebra of the gauge group is generated by the FI
(\ref{FI1}). Formally, we can think of the structure constants
$f^{ab}{}_{cd}$ as ordinary matrix elements by defining
$(f^b{}_c)^a{}_d\equiv{}f^{ab}{}_{cd}$. In other words, $f^b{}_c$
is the matrix, while $(f^b{}_c)^a{}_d$ are its matrix elements.
Then the Fundamental Identity,
\begin{equation}
f^{ab}{}_{cd}f^{de}{}_{fg} -f^{ae}{}_{fd}f^{db}{}_{cg}+
f^{eb}{}_{cd}f^{da}{}_{gf} -f^{ea}{}_{gd}f^{db}{}_{cf}=0,
\end{equation}
can be written as a commutator of two matrices:
\begin{equation}
[f^b{}_c ,
f^e{}_f]^a{}_g = f^{db}{}_{cf}(f^{e}{}_{d})^a{}_g
-f^{eb}{}_{cd}(f^{d}{}_{f})^a{}_g.
\end{equation}
It is in this sense the FI generates an ordinary Lie 2-algebra.
Specifying $f^{ab}{}_{cd}$ amounts to choosing a particular set of
structure constants of the ordinary Lie 2-algebra. In this way, the
Lie 2-algebra of the gauge group is completely determined through
the FI. It turns out that the 3-algebra structure constants
$f^{ab}{}_{cd}$ given by Eq. (\ref{SyplStruc2}) precisely generate
the Lie 2-algebra of the $Sp(2N)\times U(1)$ gauge group through the
FI.

Substituting the expression of the structure constants
(\ref{SyplStruc2}) into (\ref{susy}), the supersymmetry
transformation law now reads
\begin{eqnarray}\label{SyplSusyTrasf}
\nonumber  \delta Z^A_d &=& i\bar\epsilon^{AB}\psi_{Bd} \\
 \nonumber
 \delta \bar{Z}_{A}^{d} &=& i\bar\epsilon_{AB}\psi^{Bd} \\
 \nonumber
\delta \psi_{Bd} &=& \gamma^\mu D_\mu Z^A_d\epsilon_{AB} -
  kZ^C_a\omega^{ab}Z^A_b \omega_{dc}\bar{Z}_{C}^{c} \epsilon_{AB}-k
  Z^C_a \omega^{ab}Z^D_{b}\omega_{dc}\bar{Z}_{B}^{c}\epsilon_{CD}
\\ \nonumber
  &&-k\bar Z_C^a Z^C_a Z^A_d\epsilon_{AB}+k\bar Z_C^a Z^A_a
  Z^C_d\epsilon_{AB}-2k\bar Z_B^a Z^C_a Z^D_d\epsilon_{CD}
\\ \nonumber
\delta \psi^{Bd} &=& \gamma^\mu D_\mu \bar{Z}_A^d\epsilon^{AB}
- k\bar{Z}_C^a \omega_{ab}\bar{Z}_A^b \omega^{dc}Z^{C}_{c}
\epsilon^{AB} -k\bar{Z}_C^a \omega_{ab}\bar{Z}_D^{b}
\omega^{dc}Z^{B}_{c}\epsilon^{CD}
\\ \nonumber &&-k\bar Z_C^a Z^C_a \bar Z_A^d\epsilon^{AB}
+k\bar Z_A^a Z^C_a \bar Z_C^d\epsilon^{AB}
-2k\bar Z_C^a Z^B_a \bar Z_D^d\epsilon^{CD}\\ \nonumber
 \delta A_\mu &=&
-ik\bar\epsilon_{AB}\gamma_\mu \psi^{Ba}Z^A_a+
ik\bar\epsilon^{AB}\gamma_\mu \bar{Z}_{A}^{a}\psi_{Ba}\\
\nonumber
\delta B_\mu{}^{c}{}_d &=& ik\bar\epsilon_{AB}\gamma_\mu
\omega^{ca}Z^A_a\omega_{db}\psi^{Bb} -i k\bar\epsilon^{AB}\gamma_\mu
\omega_{db}\bar Z_A^b\omega^{ca}\psi_{Ba}\\
&& +ik\bar\epsilon_{AB}\gamma_\mu Z^{A}_{d}\psi^{Bc}
-ik\bar\epsilon^{AB}\gamma_\mu \bar Z_{A}^{c}\psi_{Bd}.
\end{eqnarray}
The Lagrangian (\ref{SymplLagrangian}) and the corresponding
supersymmetry transformation law (\ref{SyplSusyTrasf}) are in
agreement with the ${\cal N}=6, Sp(2M)\times O(2)$ superconformal
CSM theory derived from ordinary Lie 2-algebra in Ref.
 \cite{Hosomichi:2008jb}.

\section{Recasting $\CN=6$, $U(M)\times U(N)$ Theory into Symplectic 3-Algebra
Framework}
\label{ugroup}

The $\CN=6$, $U(M)\times U(N)$ theory was derived by BL in a
3-algebra framework \cite{Bagger08:3Alg} with a hermitian metric.
In this section, we will demonstrate that the $\CN=6$, $U(M)\times
U(N)$ theory can be actually recast into our symplectic
3-algebraic framework. \footnote{This section is inspired by the
referee's comment.}

To generate the direct group $U(M)\times U(N)$, we decompose the
3-algebra index $I$ into $a\a$, where $\a=1,2$, and $a=1,\cdots, MN$
stands for the bi-fundamental index of $U(M)\times
U(N)$.\footnote{In section \ref{3AlgCMS}, the index $a=1\cdots 2N$
is an $Sp(2N)$ index. We hope this does not cause any confusion.} We
then decompose the anti-symmetric metric $\omega^{IJ}$ as
\begin{equation}\label{antim}
\omega^{IJ}=\begin{pmatrix}0 & -\d^a{}_b\\ \d_a{}^b &
0\end{pmatrix}.
\end{equation}
The component formalism of the above equation reads
\begin{equation}\label{AntiMetr2}
\omega^{IJ}=\omega^{a\alpha,b\beta}
=-\delta^a{}_b\delta_{1\alpha}\delta_{2\beta}
+\delta_a{}^b\delta_{2\alpha}\delta_{1\beta}.
\end{equation}
We then decompose the structure constants $f_{IKLJ}$ as
\begin{eqnarray}\label{n6str}\nonumber
f_{IKLJ}=f_{a\alpha,c\gamma,d\delta,b\beta}
&=&f^{ac}{}_{db}\delta_{2\alpha}\delta_{1\beta}\delta_{2\gamma}\delta_{1\delta}
+f^{*ac}{}_{db}\delta_{1\alpha}\delta_{2\beta}\delta_{1\gamma}\delta_{2\delta}\\
&&+f^{bc}{}_{da}\delta_{1\alpha}\delta_{2\beta}\delta_{2\gamma}\delta_{1\delta}
+f^{*bc}{}_{da}\delta_{2\alpha}\delta_{1\beta}\delta_{1\gamma}\delta_{2\delta},
\end{eqnarray}
One still need to impose an additional constraint condition for
closing the $\CN=6$ superalgebra:
\e\label{AS3} f^{ab}{}_{cd}=-f^{ba}{}_{cd}. \ee
With these decompositions, the reality condition (\ref{HermiCondi})
becomes
\begin{equation}\label{real3}
f^{*ac}{}_{db}=f^{db}{}_{ac},
\end{equation}
and the FI (\ref{FFI}) reads
\begin{equation}\label{FI3}
f^{ab}{}_{cd}f^{de}{}_{gf} +f^{ba}{}_{fd}f^{de}{}_{gc}-
f^{ae}{}_{gd}f^{db}{}_{cf} -f^{be}{}_{gd}f^{da}{}_{fc}=0.
\end{equation}

Using (\ref{real3}), it is not difficult to verify that the RHS of
(\ref{n6str}) satisfies the desired symmetry properties. The above
three equations (\ref{AS3}), (\ref{real3}) and (\ref{FI3}) take
exactly the same forms as Eqs (\ref{SymmeOfF5}), (\ref{HermiCondi3})
and (\ref{FI}), respectively. They also take exactly the same forms
as that of BL \cite{Bagger08:3Alg}. In other words, they must belong
to the hermitian 3-algebra.

With the decomposition (\ref{antim}), it is natural to decompose a
3-algebra valued field $X_I$ as
\begin{equation}\label{6field}
X_I=\begin{pmatrix} \bar X^a\\X_a\end{pmatrix},
\end{equation}
where $\bar X^a=X^*_a$ is the complex conjugate of $X_a$. Later we
will see, with the decomposition (\ref{6field}), the complex
conjugate $X^*_{I}$ transforms in the same way as $\omega^{IJ} X_J$
(see Eq. (\ref{6global})).
We then decompose the parameter $\Lambda^K{}_L$ in the global
transformation (\ref{GlobalSym}) as
\begin{equation}\label{6parameter}
\Lambda^K{}_L=\frac{1}{2}\begin{pmatrix} -\Lambda^c{}_d & 0\\
0& \Lambda^d{}_c \end{pmatrix}.
\end{equation}
Here we require that $\Lambda^c{}_d$ is anti-hermitian, i.e.,
$\Lambda^{*c}{}_d=-\Lambda^d{}_c$, guaranteeing the anti-hermitian
condition $\Lambda^{*K}{}_L=-\Lambda^L{}_K$. With the decompositions
(\ref{antim}), (\ref{n6str}) and (\ref{6parameter}), the global
transformation (\ref{GlobalSym}) becomes
\begin{equation}\label{6global}
\begin{pmatrix} \d_{\tilde\Lambda}\bar X^a
\\ \d_{\tilde\Lambda}X_a\end{pmatrix}=
\begin{pmatrix} \Lambda^c{}_df^{ad}{}_{cb} & 0\\
0 & -\Lambda^d{}_cf^{bd}{}_{ca}\end{pmatrix}\begin{pmatrix}\bar
X^b\\X_b\end{pmatrix}=\begin{pmatrix} \tilde\Lambda^a{}_b & 0\\
0 &  -\tilde\Lambda^b{}_a\end{pmatrix}\begin{pmatrix} \bar X^b
\\X_b\end{pmatrix}=-\tilde\Lambda^{I}{}_JX_I,
\end{equation}
and Eq. (\ref{ConstrOnf}) is satisfied. Note that the reality
conditions $\Lambda^{*c}{}_d=-\Lambda^d{}_c$ and (\ref{real3}) imply
that $\tilde\Lambda^{*a}{}_b=-\tilde\Lambda^b{}_a$. So from ordinary
Lie group point of view, the matrices $-\tilde\Lambda^{I}{}_J$ are
in the $R\oplus R^*$ representation. To construct an $\CN=6$ theory,
we need only to focus on the global transformation
\begin{equation}\label{Rglobal}
\d_{\tilde\Lambda}X_a=-\tilde\Lambda^b{}_aX_b.
\end{equation}
One can easily obtain its complex conjugate. To gauge the above
symmetry, we introduce the following gauge field
\e\label{6gauge} \tilde
A_\mu{}^{a}{}_{d}=A_\mu{}^{c}{}_{b}f^{ab}{}_{cd}.\ee
Here we require that the 3-algebra tensor $A_\mu{}^c{}_b$ satisfies
the reality condition $A_\mu{}^{*c}{}_b=-A_\mu{}^b{}_c$. Eqs.
(\ref{Rglobal}) and (\ref{6gauge}) suggest us to define the scalar
and fermion fields as follows
\e Z^A_{c} \quad {\rm and}\quad\psi_{Ac},\ee
where $A$ is an $SU(4)$ R-symmetry index. Their complex conjugates
are denoted as $Z^{*A}_{c}= \BZ^c_A$ and $\psi^*_{Ac}=\p^{Ac}$. Note
that all the fields also take exactly the same forms as those in
section \ref{ClsSUSY}, though here the index $a$ runs from 1 to
$MN$, and the fields do not carry the $U(1)$ gauge group index $+$.

So, if we also follow BL's strategy to construct an $\CN=6$ theory,
the supersymmetry transformations must take exactly the same forms
as (\ref{susy}), i.e.
\begin{eqnarray}\label{susy2}
\nonumber  \delta Z^A_d &=& i\bar\epsilon^{AB}\psi_{Bd}, \\
 \nonumber
\delta \psi_{Bd} &=& \gamma^\mu D_\mu Z^A_d\epsilon_{AB} +
  f^{ab}{}_{cd}Z^C_aZ^A_b \bar{Z}_{C}^{c} \epsilon_{AB}+f^{ab}{}_{cd}
  Z^C_a Z^D_{b} \bar{Z}_{B}^{c}\epsilon_{CD}, \\
 \delta \tilde A_\mu{}^c{}_d &=&
-i\bar\epsilon_{AB}\gamma_\mu Z^A_a\psi^{Bb} f^{ca}{}_{bd} +
i\bar\epsilon^{AB}\gamma_\mu \bar{Z}_{A}^{a}\psi_{Bb}f^{cb}{}_{ad},
\end{eqnarray}
and their complex conjugates. And the equations of motion, required
by the on-shell closure of the superalgebra, must also take exactly
the same forms as those in section \ref{ClsSUSY}. Hence the
Lagrangian of this section must also take the exact form as
(\ref{Lagrangian}). Eqs (\ref{susy}) and (\ref{Lagrangian}) take
exactly the same forms as BL's general $\CN=6$ supersymmetry
transformations and the Lagrangian \cite{Bagger08:3Alg},
respectively. To derive the $U(M)\times U(N)$ theory, one just needs
to adopt the specified 3-bracket
\begin{equation}\label{UMUN}
[X,Y;\bar Z]=k(XZ^\dag Y-YZ^\dag X)
\end{equation}
from Ref. \cite{Bagger08:3Alg}, and write the hermitian bilinear
form as
\begin{equation}\label{trace}
\bar{X}^aY_a=X^*_aY_{a}=(X^{\hat n}_{m})^*Y^{\hat n}_{m}=
(X^\dag)^m_{\hat{n}} Y_{m}^{\hat{n}}={\rm tr}(X^\dag Y).
\end{equation}
Here $X^{\hat n}_{m}$ is an $m\times \hat n$ matrix, where
$m=1,\cdots, M$ is a fundamental index of $U(M)$ and $\hat
n=1,\cdots, N$ is an anti-fundamental index of $U(N)$, and $X^\dag$
is the hermitian conjugate of the $m\times \hat n$ matrix $X$.
Substituting (\ref{UMUN}) and (\ref{trace}) into (\ref{Lagrangian})
reproduces the $U(M)\times U(N)$ theory \cite{Bagger08:3Alg}.

In this way, one can `convert' Bagger and Lambert's framework into
our symplectic 3-algebraic framework, by introducing the
antisymmetric metric (\ref{AntiMetr2}) and the `map' (\ref{n6str}).

Finally, we would like to add one comment on the constraint
condition $f^{ab}{}_{cd}=-f^{ba}{}_{cd}$ (see (\ref{SymmeOfF}) or
(\ref{SymmeOfF5}), and (\ref{AS3})). This condition can be
understood as
\e\label{constraint} f_{(IJK)L}=0 .\ee
Since $f_{IJKL}=f_{IKJL}$ (see Eq. (\ref{SymofF})) and Eq.
(\ref{ConstrOnf}) implies that $f_{IJKL}=f_{LJKI}$, the above
equation is equivalent to $f_{(IJKL)}=0$. Specifically, in the
$Sp(2N)\times U(1)$ case, it becomes
\begin{equation}
f_{a-,b-,c+,d+}+f_{b-,a-,c+,d+}+f_{c+,a-,b-,d+}=0.
\end{equation}
However, since $f_{c+,a-,b-,d+}=0$ (see Eq. (\ref{SyplStruc1}) and
the explanation below (\ref{SyplStruc1})), we obtain
\begin{equation}
f_{a-,b-,c+,d+}+f_{b-,a-,c+,d+}=0,
\end{equation} which is nothing but
(\ref{SymmeOfF}). In the $U(M)\times U(N)$ case, with the `map'
(\ref{n6str}) and the reality condition (\ref{real3}), Eq.
(\ref{constraint}) becomes (\ref{AS3}). The ordinary Lie algebra
counterpart of (\ref{constraint}), first discovered in Ref.
\cite{GaWi}, is the key requirement for enhancing the $\CN=1$
supersymmetry to $\CN=4$.
\section{Conclusions and Discussions}
\label{Conclusion}

In this paper, we first introduce the notion of symplectic
3-algebras. We then give a formulation of ${\cal N}=6$
superconformal Chern-Simons-matter (CSM) theory with $SU(4)$
R-symmetry based on the symplectic 3-algebras. By specifying the
3-brackets, we derive the ${\cal N}=6, Sp(2N)\times U(1)$
superconformal CSM theory in our framework. We also recast the
${\cal N}=6, U(M)\times U(N)$ into the symplectic 3-algebraic
framework.

The ${\cal N}=6$ superconformal CSM theories in three dimensions
have been completely classified in Ref. \cite{Schn} by using group
theory. The ${\cal N}=6$ CSM theories can also be classified by
super Lie algebras \cite{GaWi,Hosomichi:2008jb,Kac}. Essentially,
only two types are allowed: with gauge group $Sp(2N)\times U(1)$ and
$U(M)\times U(N)$, respectively. Therefore our approach provides a
unified 3-algebra framework to describe all known ${\cal N}=6$
superconformal theories. Though our approach to the $\CN=6$ theories
is essentially equivalent to that of BL \cite{Bagger08:3Alg}, our
formulation is slightly different from the latter, in that ours is
more suited to the case with gauge group $Sp(2N)\times U(1)$.

The question of reformulating the known CMS models in a 3-algebra
approach is not merely of mathemtical interests. More important is
whether or not the M2-branes physics would become more transparent
if looked through a new mathematical framework such as 3-algebras.

It would be also nice to find the gravity dual of the ${\cal N}=6,
Sp(2N)\times U(1)$ CSM theory, and to investigate the integrability
from both the gauge theory side and string/M theory side
\cite{Minahan}. It would be interesting to generalize the symplectic
3-algebra model so that its gauge group has more general product
structure, like those in quiver gauge theories.

\section{Acknowledgement}
We thank the referee for his/her comment. The work of YSW is
partially supported by US NSF grant No. PHY-0756958 and by a grant
from the Foundational Questions Institute.

\appendix

\section{Conventions and Useful Identities}\label{Identities}

In $1+2$ dimensions, the gamma matrices are defined as
$\{\gamma_\mu, \gamma_\nu\}= 2\eta_{\mu\nu}$. For the metric we use
the $(-,+,+)$ convention. The gamma matrices can be defined as the
Pauli matrices: $\gamma_\mu=(i\sigma_2, \sigma_1, \sigma_3)$,
satisfying the important identity
\begin{equation}
\gamma_\mu\gamma_\nu=\eta_{\mu\nu}+\varepsilon_{\mu\nu\lambda}\gamma^{\lambda}.
\end{equation}
We also define
$\varepsilon^{\mu\nu\lambda}=-\varepsilon_{\mu\nu\lambda}$. So
$\varepsilon_{\mu\nu\lambda}\varepsilon^{\rho\nu\lambda} =
-2\delta_\mu{}^\rho$.

The following identities are adopted from Ref. \cite{Bagger08:3Alg}.
In $1+2$ dimensions the Fierz transformation is
\begin{equation}
(\bar\lambda\chi)\psi = -\frac{1}{2}(\bar\lambda\psi)\chi
-\frac{1}{2} (\bar\lambda\gamma_\nu\psi)\gamma^\nu\chi.
\end{equation}
Some useful $SU(4)$ identities are
\begin{eqnarray}
\nonumber
\frac{1}{2}\bar\epsilon^{CD}_1\gamma_\nu\epsilon_{2CD}\,\delta^A_B
&=&\bar\epsilon^{AC}_1\gamma_\nu\epsilon_{2BC}
-\bar\epsilon^{AC}_2\gamma_\nu\epsilon_{1BC}\\
 \nonumber
2\bar\epsilon^{AC}_1\epsilon_{2BD}
-2\bar\epsilon^{AC}_2\epsilon_{1BD}
&=&\bar\epsilon^{CE}_1\epsilon_{2DE}\delta^A_B
-\bar\epsilon^{CE}_2\epsilon_{1DE}\delta^A_B\\
\nonumber
&-&\bar\epsilon^{AE}_1\epsilon_{2DE}\delta^C_B
+\bar\epsilon^{AE}_2\epsilon_{1DE}\delta^C_B\\
&+& \bar\epsilon^{AE}_1\epsilon_{2BE}\delta^C_D
-\bar\epsilon^{AE}_2\epsilon_{1BE}\delta^C_D\\
\nonumber
&-& \bar\epsilon^{CE}_1\epsilon_{2BE}\delta^A_D
+\bar\epsilon^{CE}_2\epsilon_{1BE}\delta^A_D\\
\nonumber \frac{1}{2}\varepsilon_{ABCD}
\bar\epsilon^{EF}_1\gamma_\mu\epsilon_{2EF}
&=&\bar\epsilon_{1AB}\gamma_\mu\epsilon_{2CD}
-\bar\epsilon_{2AB}\gamma_\mu\epsilon_{1CD}\\
&+& \bar\epsilon_{1AD}\gamma_\mu\epsilon_{2BC}
-\bar\epsilon_{2AD}\gamma_\mu\epsilon_{1BC}\\
\nonumber
&-&\bar\epsilon_{1BD}\gamma_\mu\epsilon_{2AC}
+\bar\epsilon_{2BD}\gamma_\mu\epsilon_{1AC}.
\end{eqnarray}


\begin{thebibliography}{99}

\bibitem{Schwarz2004} J. Schwarz, "Superconformal Chern-Simons
Theories," JHEP, 0411:078 (2004); arXiv:hep-th/0411077.

\bibitem{gaiottoyin}
  D.~Gaiotto and X.~Yin,
  ``Notes on superconformal Chern-Simons-matter theories,''
  JHEP {\bf 0708}, 056 (2007)
  [arXiv:0704.3740 [hep-th]].

\bibitem{CSW1} W. Chen, G.W. Semenoff and Y.S. Wu, ``Scale
and Conformal Invariance in Chern-Simons-Matter Field Theory'',
Phys. Rev. D44, 1625 (1991).

\bibitem{CSW0} W. Chen, G. Semenoff and Y.S. Wu, ``Probing Topological
Features in Perturbative Chern-Simons Gauge Theory'', Mod. Phys. Lett.
A5, 1833 (1990).

\bibitem{CSW2} W. Chen, G.W. Semenoff and Y.S. Wu, ``Two-Loop Analysis
of Chern-Simons-Matter Theory'', Phys. Rev. D46, 5521 (1992);
arXiv:hep-th/9209005.

\bibitem{Piguet} O.M. Del Cima, D.H.T. Franco, J.A. Helayel-Neto and
O. Piguet, ``An algebraic proof on the finiteness of
Yang-Mills-Chern-Simons theory in D=3'', Lett. Math. Phys. 47, 265
(1999); arXiv:math-ph/9904030.

\bibitem{Nambu} Y. Nambu, Generalized Hamiltonian mechanics, Phys.
Rev. \textbf{D7} (1973), 2405-2412.

\bibitem{Limiao:1999fm} H. Awata, M. Li, D. Minic, T. Yoneya,
 ``On the quantization of nambu brackets,''
arXiv:hep-th/9906248.

\bibitem{Bagger}
  J.~Bagger and N.~Lambert,
  ``Modeling multiple M2's,''
  Phys.\ Rev.\  D {\bf 75}, 045020 (2007),
  arXiv:hep-th/0611108;
``Gauge Symmetry and Supersymmetry of Multiple M2-Branes,''
  Phys.\ Rev.\  D {\bf 77}, 065008 (2008),
  arXiv:0711.0955 [hep-th];
  J.~Bagger and N.~Lambert,
  ``Comments On Multiple M2-branes,''
  JHEP {\bf 0802}, 105 (2008),
  arXiv:0712.3738 [hep-th].


\bibitem{Gustavsson}
  A.~Gustavsson,
  ``Algebraic structures on parallel M2-branes,''
  arXiv:0709.1260 [hep-th];
``Selfdual strings and loop space Nahm equations,''
  JHEP {\bf 0804}, 083 (2008),
  arXiv:0802.3456 [hep-th].

\bibitem{DMPV}
  J.~Distler, S.~Mukhi, C.~Papageorgakis and M.~Van Raamsdonk,
  ``M2-branes on M-folds,''
  JHEP {\bf 0805}, 038 (2008)
  arXiv:0804.1256 [hep-th].


\bibitem{LambertTong}
  N.~Lambert and D.~Tong,
  ``Membranes on an Orbifold,''
  arXiv:0804.1114 [hep-th].

\bibitem{Gauntlett}
Jerome P. Gauntlett, Jan B. Gutowski, ``Constraining Maximally
Supersymmetric Membrane Actions,'' arXiv:0804.3078 [hep-th].

\bibitem{Papadopoulos}
G. Papadopoulos ``M2-branes, 3-Lie Algebras and Plucker relations ,"
arXiv:0804.2662 [hep-th].

\bibitem{Lorentzian3Alg}
  J.~Gomis, G.~Milanesi and J.~G.~Russo,
  ``Bagger-Lambert Theory for General Lie Algebras,''
  arXiv:0805.1012 [hep-th];
  S.~Benvenuti, D.~Rodriguez-Gomez, E.~Tonni and H.~Verlinde,
  ``N=8 superconformal gauge theories and M2 branes,''
  arXiv:0805.1087 [hep-th];
  P.~M.~Ho, Y.~Imamura and Y.~Matsuo,
  ``M2 to D2 revisited,''
  arXiv:0805.1202 [hep-th];

\bibitem{GhostFree}
  M.~A.~Bandres, A.~E.~Lipstein and J.~H.~Schwarz,
  ``Ghost-Free Superconformal Action for Multiple M2-Branes,''
  arXiv:0806.0054 [hep-th];

\bibitem{CalN8SYM}
  J.~Gomis, D.~Rodriguez-Gomez, M.~Van Raamsdonk and H.~Verlinde,
  ``Supersymmetric Yang-Mills Theory From Lorentzian Three-Algebras,''
  arXiv:0806.0738 [hep-th].

\bibitem{ABJM}
O.~Aharony, O.~Bergman, D.~L.~Jafferis and J.~Maldacena,
  ``N=6 superconformal Chern-Simons-matter theories, M2-branes and their
  gravity duals,''
  arXiv:0806.1218 [hep-th].

\bibitem{Benna}
  M.~Benna, I.~Klebanov, T.~Klose and M.~Smedback,
  ``Superconformal Chern-Simons Theories and $AdS_4/CFT_3$ Correspondence,''
  arXiv:0806.1519 [hep-th].

\bibitem{Schwarz}
  M.~Bandres, A.~Lipstein and J.~Schwarz
 ``Studies of the ABJM Theory in Formulation with Manifest $SU(4)$ R-Symmetry,''
  arXiv:0807.0880 [hep-th].

\bibitem{Bergshoeff:2008cz}
  E.~A.~Bergshoeff, M.~de Roo and O.~Hohm,
  ``Multiple M2-branes and the Embedding Tensor,''
  arXiv:0804.2201 [hep-th];
  E.~A.~Bergshoeff, M.~de Roo, O.~Hohm and D.~Roest,
  ``Multiple Membranes from Gauged Supergravity,''
  arXiv:0806.2584 [hep-th].

\bibitem{Bergshoeff}
  E.~A.~Bergshoeff, O.~Hohm, D.~ Roest, H.~Samtleben and  E.~Sezgin,
  ``The Superconformal Gaugings in Three Dimensions,''
  arXiv:0807.2841 [hep-th].

\bibitem{GaWi}
  D.~Gaiotto and E.~Witten,
   ``Janus Configurations, Chern-Simons Couplings, And The Theta-Angle in N=4
  Super Yang-Mills Theory,''
  arXiv:0804.2907 [hep-th].

\bibitem{HosomichiJD}
  K.~Hosomichi, K.~M.~Lee, S.~Lee, S.~Lee and J.~Park,
  ``N=4 Superconformal Chern-Simons Theories with Hyper and Twisted Hyper
  Multiplets,''
  arXiv:0805.3662 [hep-th].

\bibitem{Hosomichi:2008jb}
  K.~Hosomichi, K.~M.~Lee, S.~Lee, S.~Lee and J.~Park,
 ``N=5,6 Superconformal Chern-Simons Theories and M2-branes on Orbifolds,''
  arXiv:0806.4977 [hep-th].

\bibitem{Aharony:2008gk}
  O.~Aharony, O.~Bergman and D.~L.~Jafferis,
  ``Fractional M2-branes,''
  arXiv:0807.4924 [hep-th].

\bibitem{Bagger08:3Alg}
  J.~Bagger and N.~Lambert,
  ``Three-Algebras and N=6 Chern-Simons Gauge Theories,''
  arXiv:0807.0163 [hep-th].

\bibitem{Schn}
  M.~ Schnabl and Y.~Tachikawa
 ``Classifiction of ${\cal N}=6$ superconformal theories of ABJM type,''
  arXiv:0807.1102 [hep-th].

\bibitem{Kac}
V.~G.~Kac, ``Lie Superalgebras,'' Adv. Math. \textbf{26} (1977) 8.

\bibitem{Minahan}
C.~ Krishnan, C.~ Maccaferri, ``Membranes on Calibrations," JHEP
{\bf 0807}, 005 (2008), arXiv:0805.3125[hep-th];
C.~ Krishnan, ``$AdS_4/CFT_3$ at One Loop," JHEP {\bf 0809}, 092
(2008), arXiv:0807.4561 [hep-th];
L.~ F.~ Alday.~ G.~ Arutyunov, D.~ Bykov , ``Semiclassical
Quantization of Spinning Strings in $AdS_4 \times CP^3$," JHEP {\bf
0811}, 089 (2008), arXiv:0807.4400 [hep-th];
T.~ McLoughlin, R.~ Roiban, ``Spinning strings at one-loop in $
AdS_4 \times P^3 $," JHEP {\bf 0812}, 101 (2008), arXiv:0807.3965
[hep-th];
 J.~Minahan, W.~Schulgin, and K.~Zarembo, `` Two
loop integrability for Chern-Simons theories with ${\cal N}=6$
supersymmetry," arXiv:0901.1142 [hep-th].

\end{thebibliography}
\end{document}